# SMARTAJWEED
## AUTOMATIC RECOGNITION OF ARABIC QURANIC RECITATION RULES

Ali M. Alagrami[1] and Maged M. Eljazzar[2]

[1]Department of Computer Science, University of Venice, Italy
[2]Faculty of Engineering, Cairo University, Egypt

*ABSTRACT*

*Tajweed is a set of rules to read the Quran in a correct Pronunciation of the letters with all its Qualities, while Reciting the Quran. which means you have to give every letter in the Quran its due of characteristics and apply it to this particular letter in this specific situation while reading, which may differ in other times. These characteristics include melodic rules, like where to stop and for how long, when to merge two letters in pronunciation or when to stretch some, or even when to put more strength on some letters over other. Most of the papers focus mainly on the main recitation rules and the pronunciation but not (Ahkam AL Tajweed) which give different rhythm and different melody to the pronunciation with every different rule of (Tajweed). Which is also considered very important and essential in Reading the Quran as it can give different meanings to the words. In this paper we discuss in detail full system for automatic recognition of Quran Recitation Rules (Tajweed) by using support vector machine and threshold scoring system.*

## 1. INTRODUCTION

The correct pronunciation of Al-Quran is based on the "Tajweed rules" which are essential to follow while recitation of the Al-Quran [1]. "Jawwada" is the root word from which the Arabic word Tajweed is derived. Jawwada means to make better or improve the pronunciation and it's accuracy during recitation [2].

Tajweed is obligatory for all Muslims as following Tajweed rules properly: it becomes easy to read the Al-Quran accurately [3-4]. Because if any mistake happened in Tajweed's rules during the recitation of Al-Quran, it can alter the definite meaning of that word, which is mispronounced. Many of the Islamic education studies showed that reciting The Al-Quran has always been a complex issue. First studies demonstrate that the students face many problems to use the correct Tajweed rules during reciting and reading Al-Quran [5-7]. Secondly, many Muslims are non-Arabic native speakers. At last, non-native Arabic speakers may face issue understand the difference in meaning for the same words; as low regularity in the reciting of Al-Quran is an issue that majorly contributes to the misreading or not properly reciting the Al-Quran among Muslims [8].

One of the traditional methods in learning Tajweed rules is "Talaqqi Musyafahah's" [9] and like any traditional learning process, it is supervised by the teacher, who you may not be completely sure about his skills and abilities. on the other hand, problems as the teaching environment, lack of attention, or even surroundings noises which may contribute to the possibility of non-accurate hearing and prone to errors. Moreover, the hours given for learning are also limited. Currently, several applications utilize smart phones to simplify the Tajweed learning process. In a previous work, we introduced Imam [24], an application to learn Al-Quran with gamification, and make it simplify the Tajweed learning process between Sheikh and student.

Based on the problem mentioned above, it's been observed that it is necessary to advance the existing way of Tajweed learning. In this paper we propose a system capable of automatic recognition of four different rules of Tajweed with very high Accuracy 99% which is very promising results for further improvements, also the whole system is implemented in IMAM Application in its beta version.

## 2. LITERATURE REVIEW

Computer-Aided Pronunciation Learning (CAPL) has been presented and attained special consideration in past years. In second language teaching, several research works have been made to develop these systems [15, 16]. After facing many challenging issues and complications in knowing Quranic letters, specifically in checking the Tajweed's rules, the researchers took many serious actions in these past years.

While researching on the Quranic verse recitation recognition, a group of study headed by H.Tabbal, et al. (2006) [17] observed the delimitation system of Quranic verse in the form of audio files via this speech recognition system. This project's approach concerns more with the computerized delimiter that can excerpt verse throught audio files. Examination techniques for every phase were elaborated and assessed using several reciters techniques, which recite surah "Al-Ikhlas". The most significant rules of Tajweed and tarteel were also the part of the discussion that can Impact the recognition of a particular recitation in Surah Al-Ikhlas.

To assist the learner in order to recite Al-Quran through a collaborative learning method, computerized Tajweed Inspection Rules Engine for the Learning the quran have been presented [18]. This proposed system will helps the students to recite the Al-Quran in the absence of any instructor. The recitation done by the user will get processed by using the system, and revision of recitation is done. Still, the system's engine can only be verified based on certain Tajweed rules of surah Al-Fatihah.

The system of E-Hafiz functions similarly to a Hafiz that can help in the recitation of learning Al-Quran by reducing errors and inaccuracies during practice of the recitation. The Hafiz is a proficient person who performs functions similar to a teacher, i.e., first to listen and then teach Tajweed's correct recitation [19]. But still, this model concerns to only those who are already familiar with the Tajweed. In offline mode, the system does works. This system does not point any of the user's mistakes concurrently during recitation [13].

With the use of speech recognition technology, another Tajweed rule checking has been recently purposed to assist students in learning and reviewing correct recitation of Al-Quran by on their own [3]. This system can identify and highlight the incompatibility among the students recitations with that of the experienced and expert instructors stored in the database. The feature classification system has implemented the HMM algorithm, and for feature extraction, the system adopted the MFCC algorithm.

Moreover, another study using the image processing technique based on Tajweed's automatic rules is proposed [20]. The method of this research is constrained only by the laws of Idgham.

Next, input image passed via process of pre-processing, consisting of four sub processes: binary conversion, thinning and flip, grayscale conversion, and the word segmentation. Hereafter, six characteristics of shape descriptors were extracted from each input image: minor axis length, major axis length, filled area, eccentricity, perimeter and solidity. To understand the two forms of Idgham Laws that are Idgham Maal Ghunnah and Idgham Bila Ghunnah, a method of k-Nearest Neighbor (k-NN) is used. To assess the proposed analysis's efficiency, 180 test images were analyzed, which demonstrated the classification accuracy of 84.44%. The research outcome is supposed to instantly understand Tajweed's rules and allow the speakers to correctly recite the Al-Quran.

## 3. PROPOSED APPROACH

Our main objective in this paper is to build a complete system capable of recognizing the different rules of Tajweed in an audio. And determine whether it was pronounced correctly or not in a percentile metric. moreover, build an application over this system. which will enable hundreds of millions of Muslims, and all non arabic speakers all around the globe to learn how to read the Holy Quran Correctly. In our System we considered four different rules (Edgham Meem, Ekhfaa Meem, takhfeef Lam, Tarqeeq Lam) for each rule we collected a dataset from universities, Expert Volunteers and paid Experts. The dataset contains the right way for the pronunciation and different wrong ways in which it can be pronounced. taking in consideration Arabic natives and non-natives. Then we built a system that can work not only on specific Verses but a system able to recognize the rule in General that can Recognize the Rule in any verse in the holy Quran. In the first section we will discuss the overall system pipeline after that in the second section we will talk more about the Feature Extraction methodology and the Machine learning technique and at last we will show the results and conclusion and the future work.

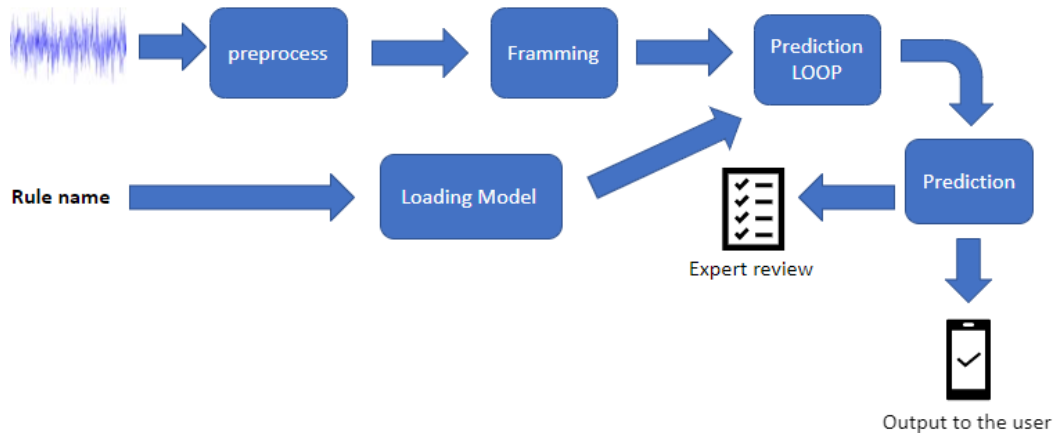

Figure 1. Tajweed Abstracted System Architecture

### 3.1. Abstract System Design.

As mentioned earlier the system main goal is to recognize the different rules of "Tajweed" and to indicate whether it's correct or not with a focused objective to make it easier for all Muslims whatever their native language to learn how to read the Quran probably. The input to the system is composed of two main Parameters, first the audio file which is supposed to contain the "Tajweed" rule, along with the name of the rule which should be found in the audio file. By knowing the name of the rule, we load the model built specifically for this rule and send it to the prediction loop (Figure 1). At the same time the input audio is preprocessed, framed into 4 seconds with stride 0.5 seconds. The frames and the loaded model are sent to the prediction loop which loops over the frames and uses the model loaded to predict the part of the audio which contains the rule. The prediction on each frame is compared to a set of specific thresholds for each recognition (Right/Wrong) for this specific rule. if the prediction is higher than that threshold the prediction will be taken in consideration otherwise it will be discarded.

The prediction loop sends the list of the predictions to the prediction module which compares the predictions and chooses the one to be shown to the user based on a certain rule. (Right pronunciation/Wrong Pronunciation and how close is the user from the right pronunciation) Moreover it saves all the records and the prediction which is then sent to an expert to review correctly so as to be used later to refine the model and the system periodically.

### 3.2. Dataset

First, we used a limited dataset we have to build the beta system. The dataset contained about roughly 80 record for each Rule name and type with a total of 657 recordings of 4 different rules each has two different types (4 rules, 8 classes). To Collect more dataset and refine the model we launched beta feature in our application by which we received about 600 more records from different app users in only 2 months and more is coming every week with 30+ new users ever month. which will help us a lot in refining the Model. all the recordings we receive is reviewed and labelled by an expert.

### 3.3. Processing

before this stage each audio file in the dataset is manually cut so as to contain only the part in which the rule is pronounced (Right or Wrong), all the audios sampling rate was changed to 8000 Hz, then we took the average length for all the audio files which was about 4 seconds , then we converted the length of all the other audio files to be 4 seconds. if it is less than 4 seconds we add silence, if longer randomly truncate is done.

### 3.4. Feature Extraction

Our approach employs filter banks as a baseline method for feature extraction. Filter banks were motivated by the nature of the speech signal and the human perception of such signals[21]. First the signals are cut into frames each of size 25 ms with stride 10 ms. hamming window is applied for minimizing the disruptions at the starting and at the end of each frame , then we perform FFT to obtain the magnitude frequency response to get the spectrum which is then subjected to 70 triangular band pass filters in order to get smooth spectrum, and to reduce the size of the features involved.

### 3.5. Classification Model

For the Classification task we used Support Vector Machine (SVM) which was first Introduced in 1992 by Boser, Guyon and Vapnik [22] . SVM is considered a supervised machine learning method that can be used for classification, regression, and Outliers Detection. However, it is mostly used in classification problems. SVM also belongs to the general category of Kernel methods [23] which are methods that depend on the Data through only the dot products. kernels use this fact and introduce this dot product in a possibly higher dimensional feature space which in turns make it easier to find nonlinear decision boundaries using linear classifier methods and allow us to use the classifier on data that does not have fixed dimension this trick used here is called the "kernel trick".

As mentioned The support vector machine is a binary classifier Algorithm tries to find the Maximum marginal hyperplane which is the plane with maximum distance from the two classes introduced. Let $\{x_i, y_i\}$ be sample of the dataset in which $x_i$ is an input features and $y_i$ is the corresponding class value (-1,+1) "P" and "N" . so that we have

$$y_i(w^T x_i + b) \geq 1 \ \ for\ all\ x_i$$

To make computations easier and without loss of generality the pair $(w, b)$ can be rescaled such that:

$$min_{i=1\ldots l} |(w^T x_i + b)| = 1$$

To restrict the hypothesis space, the support vector machine tries to find the simplest solution that classifies the input data correctly. The learning problem is therefore can be written as: Minimization $\|W\|^2 = W^T W$ subjected to the constraints of linear separability. This is equivalent to maximizing the margin between the convex Envelope of the two classes we have. The optimization is now a convex quadratic programming problem.

$$Minimization\ \Phi(W) = \frac{1}{2} \|W\|^2$$

$$\text{Subject to } y_i(W^T x_i + b) \geq 1, \ for\ all\ i$$

And as the input data appears only in the form of dot product $x^T x'$ the Kernel $K(x, x')$ can be introduced to map the data into higher dimensional space. Also one of the main attributes that affect the performance of the SVM is the penalty parameter "C", where lower value of "C" encourages greater margin for a lower accuracy. On the other hand, the "gamma" parameter can be seen as the inverse of the radius of influence of the data points selected by the model as support vectors.

To build the model we used a radial basis kernel and tuned the hyperparameters "C" and "gamma" through a grid search algorithm where the best performance was achieved when the "C" is tuned to 1 and "gamma" to 0.1

## 4. TESTING AND RESULTS

Each model in the system was tested against 30% of the Data recorded with a validation accuracy of 99%. Then the system as one entity is then tested against full verses to extract the rules, a sample of the testing table can be seen in (Table 2) where each row is a verse in Quran in which certain rule should be found. The system was not only tested to recognize the rule but also extract the exact timing in which the rule is recognized and how close it is to the threshold set earlier for this specific rule name and type. In (Figure 2) you can see the results of the system on two test audios one on "Edgham meem" and the other on "Tarqeeq lam" where the green line depicts the rule starting time recognized by the system while the red line is the true line set by the expert. The thresholds for each rule name and type (Right, Wrong) are selected so as to minimize the false positive as much as possible in the testing Data-set. But the thresholds are subjected to be changed in general as a response to the overall performance after deploying the system.

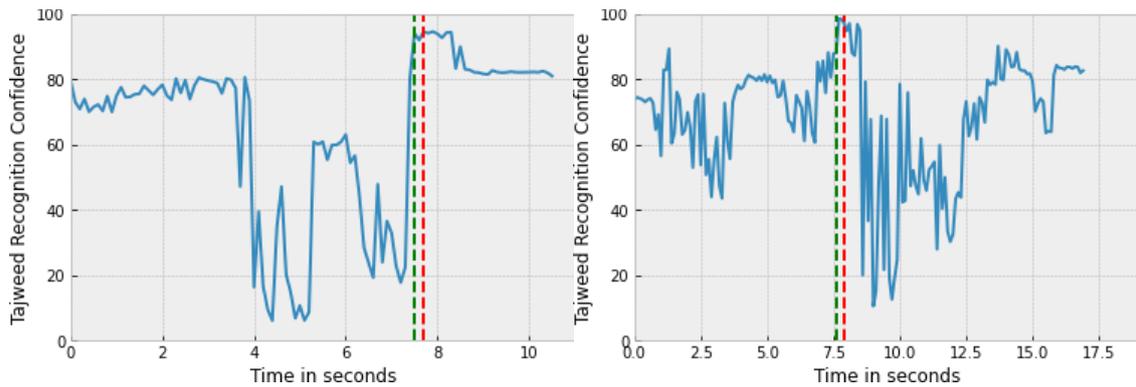

Figure 2. the figure on the left shows the output result of the model on edgham meem test sample and on the right output results on tarqeeq lam test sample

Table 1. Confusion Table for the 4 SVM models

| Rule Name | True Positive | False Positive | True Negative | False Negative |
|---|---|---|---|---|
| Edgham Meem | 30 | 0 | 30 | 0 |
| Tafkheem Lam | 30 | 0 | 30 | 0 |
| Tarqeeq Lam | 29 | 0 | 30 | 1 |
| Ekhfaa Meem | 30 | 0 | 30 | 0 |

Table 2. Samples of the verses used in the testing phase

| Rule Name (Right) | Chapter | Verse |
|---|---|---|
| Ekhfaa Meem | 41 | 52 |
| Ekhfaa Meem | 7 | 45 |
| Ekhfaa Meem | 40 | 75 |
| Ekhfaa Meem | 38 | 46 |
| Tafkheem Lam | 2 | 9 |
| Tafkheem Lam | 2 | 10 |
| Tafkheem Lam | 8 | 16 |
| Tafkheem Lam | 2 | 149 |
| Tafkheem Lam | 14 | 51 |
| Tarqeeq Lam | 2 | 67 |
| Tarqeeq Lam | 14 | 5 |
| Tarqeeq Lam | 18 | 26 |
| Tarqeeq Lam | 16 | 41 |
| Edgam Meem | 38 | 62 |
| Edgam Meem | 38 | 4 |
| Edgam Meem | 6 | 164 |
| Edgam Meem | 4 | 1 |
| Edgam Meem | 37 | 151 |
| Edgam Meem | 39 | 60 |

## 4. CONCLUSION AND FUTURE WORK

In this paper, we proposed the new approach that we used to build the Tajweed System in IMAM Application. By treating the problem as a normal binary classification problem. The input Audio Signal is first preprocessed then goes through the features extraction in which we used 70 Filter banks. And for the classification, we used SVM with the threshold scoring method which allowed us to have full control over the system rule recognition capability and allowed us to tune it to get the best results. The Work done here will be continued to include even more rules with main target goal to include all the rules for Recitation in the Quran and deploy it in our application free to use for anyone, anywhere.